\documentstyle[emulateapj,onecolfloat,psfig]{article}   

\begin{document}
\twocolumn[      

\title{Kuiper Belt Object Sizes and Distances from Occultation Observations}
\author{Asantha Cooray}
\affil{Theoretical Astrophysics, Mail Code 130-33, California Institute of Technology, Pasadena, CA 91125\\
E-mail:asante@tapir.caltech.edu}


\begin{abstract}
There are several observational campaigns under way to detect kilometer size foreground  Kuiper Belt Objects (KBOs)
through their occultation of background stars.
The interpretation of an occultation light curve, unfortunately, is
affected by a geometric degeneracy such that one is unable to determine the KBO size independent of
its distance. This degeneracy can be broken through a precise measurement
of the relative velocity obtained from simultaneous observations of individual events. 
While an array of telescopes spread over
an area of few  square kilometers can be employed, it is unlikely that the relative velocity can be
measured to the required accuracy to help break this geometric degeneracy. The presence of
diffraction fringes in KBO occultation light curves, when projected sizes of occulted stars are smaller than the
Fresnel scale, improves determination of size and distance significantly. In this regard, we discuss the potential role of 
a dedicated satellite mission for KBO occultation observations. If dwarf stars at the V-band magnitudes of 
14th and fainter can be monitored at time intervals of 0.1 seconds with normalized flux errors at the level of 1\%,  
the occultation observations will allow individual KBO sizes and distances to be determined at the level
of a few percent or better.
\end{abstract}
]

\section{Introduction}

The Kuiper Belt Object (KBO) population may provide useful clues to understand the formation and evolution of our
Solar system and many others. The presence of KBOs beyond the orbit of Neptune is now well established; current estimates suggest
 that there is a total of order $10^5$ objects with sizes greater than 100 km (see Luu
\& Jewitt 2002 for a recent review).  While sensitive photometric surveys may be capable of detecting
KBOs with sizes greater than several tens of kilometers, the size distribution of KBOs with sizes of order
a kilometer and below is useful for comparison with numerical models (Kenyon 2002). 
While imaging surveys are unlikely to detect
kilometer-sized KBOs, as the expected R-band magnitude is of order $\sim$ 30 or fainter, occultation surveys
provide a much needed probe to extract population statistics at this low end of the size distribution.
Unlike significant dependence on distance associated with reflected light, occultation observations (e.g.,
Elliot \& Olkin 1996) can also 
easily extend statistics of minor bodies beyond the current range of distances encountered.

In previous discussions, the use of occultation observations to understand the KBO population was
considered (Bailey 1976; Brown \& Webster 1997; Roques \& Moncuquet 2000; 
Cooray \& Farmer 2003).  Due to the potentially large number of KBOs at the low end of the size distribution,
the probability for occultation is at the level of $\sim$ 0.1 per star per year.
This event rate can be compared to that of galactic microlensing surveys where the 
optical depth is at the level of 10$^{-6}$ requiring monitoring
observations of order million stars or more. The KBO occultations, however,  have a significant disadvantage in
that the time duration of events is typically at the level of a few tenths of a 
second or below. This puts severe constraints on the instrumentation side for reliable detections.

In Cooray \& Farmer (2003), we explored the use of small telescopes arrays, such as the one planned by the 
Taiwanese-American Occultation Survey (TAOS; Liang et al. 2002), and
showed that these surveys  can be used to constrain the size distribution of KBOs between few 100 meters and 10 km.
Such statistical constraints assume prior knowledge on the KBO distance and velocity distributions.
While this approach may be useful for first generation surveys,
for precise knowledge on the KBO population, it is useful to constrain the KBO size distribution independently.
A step towards this direction is to extract parameters, such as the distance and size, from
individual light curves. Unfortunately, such an extraction is
affected by a geometric degeneracy between the KBO size and its distance such that one cannot determine one of these parameters 
independent of the
other. This degeneracy, however, can be partly broken with a measurement of the relative velocity associated with the event
and, in general, require observations with multiple telescopes for this purpose.

The occultation light curves are expected to show effects related to  diffraction  
when events involve background stars whose projected sizes at KBO distances are below the Fresnel scale. 
In addition to enhancing the event rate (Roques \& Moncuquet 2000; Cooray \& Farmer 2003), 
through the presence of fringes, the diffraction effect has another advantage in that
one can break the degeneracy associated with KBO size and distance with no
independent measurement of the relative velocity.  In practice,  such a parameter estimation require
flux measurements of background stars at the level of a few percent or better.
An interesting possibility for this purpose involves a dedicated satellite mission for KBO occultations.

The discussion is organized as follows. In the next Section, 
we will summarize important aspects related to KBO
occultations, such as the normalized light curve. 
In Section~3, we discuss practical considerations related to KBO occultations
and consider requirements for an array of sub-meter class telescopes based on the importance of a
relative velocity measurement. In the same Section, we also discuss the ability to extract
information from individual light curves obtained at high precision with a space-based monitoring
mission for KBO occultations. 

\begin{figure}[t]
\centerline{\psfig{file=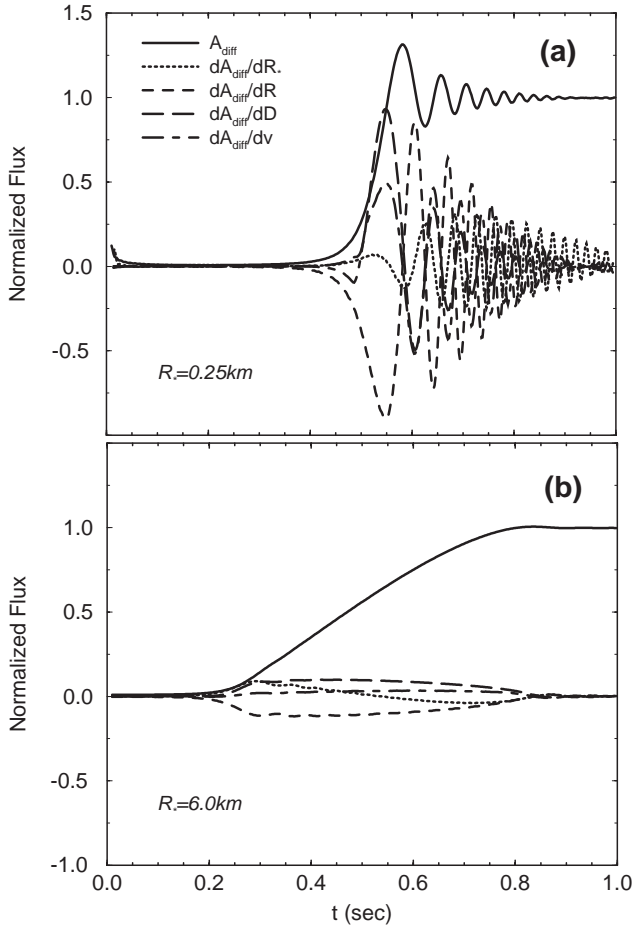,width=3.3in,angle=0}}
\caption{The fractional dimming of background stellar intensity
during an occultation event by a foreground KBO of 10 km in radius with a projected stellar radius of
0.25 km (a) and 6 km (b). Note that the full light curve is symmetric about the central
position (taken to be at t=0 seconds) and we only show one-half here.
The Fresnel scale is taken to be 0.7 km, consistent 
with the value at a distance of 40 AU and an observing wavelength of 500 nm. 
In addition to the light curve (solid line), we also show the derivatives of the light curve
in terms of the physical parameters related to the occultation. As the star size is increased, the
diffraction fringes disappear and the information contained within the derivatives decrease.
This results in a substantial lowering of the ability to extract parameters from occultations that
involve larger projected stars, which also happen to be the brighter ones, at KBO distances.}
\label{fig:diff}
\end{figure}

\section{KBO Occultations}

In general, one can characterize
an occultation of a background star by a foreground KBO with two parameters involving the
duration, $\Delta T$, of the occultation  and the fractional flux
decrease, $A$, of the background star during the occultation.  In the geometric description of an occultation,
these two observables are related via $\Delta T = 2 D \phi_* (1+\sqrt{A})/v$,
where $D$ is the distance to the foreground KBO, $v$ is the relative velocity, and $\phi_*$
is the angular size of the background star. The latter is what one obtains from
stellar physics based arguments, such as through colors or a spectrum, or from interferometric observations
(van Belle 1999). Since $A= R^2/R_*^2$, with KBO size, $R$, and projected star size, $R_*(=D\phi_*)$,
with occultation light curves, the
combination of $R/v$ and $D/v$ are well constrained when prior information related 
to the stellar angular size is available. 
A precise measurement of the relative velocity during the occultation event is needed 
to estimate $R$ and $D$ separately, otherwise a small KBO near by or a large KBO far away can be equally used to
interpret a given light curve. The relative velocity measurement 
requires observations of the same event at two separate locations that happen to lie within the
same occultation path. Since occultation
observations are mostly sensitive to kilometer size KBOs, this can be achieved with an array of 
telescopes  where individual  elements are spread over an area of few square kilometers or more. 
For comparison, a similar degeneracy exists for microlensing light curves involving the lens mass and its distance and
to help break this degeneracy, at least one additional observation is required from a satellite 
at Astronomical-Unit distances (e.g., Gould 1995).

In the case of KBO occultations, another possibility to break the degeneracy between distance and size involves 
the presence of diffraction  fringes in occultation light curves.
At KBO distances, the diffraction effect becomes important especially when
the projected star size involved with an occultation is smaller than the Fresnel scale given by
 $R_{\rm F} = \sqrt{\lambda D/2\pi}$. At $D \sim 40$ AU, $R_{\rm F} \sim 0.7$ km when $\lambda=500$ nm
and is typical of projected stellar sizes, especially for  main sequence stars with magnitudes
around 14.   With diffraction, 
the normalized flux variation as a function of the separation between the foreground KBO and the background star, $r$, is  
(Roques et al. 1987):
\begin{equation}
A_{\rm diff}(R_*,R,|{\bf r}|) = \int_0^{R_*} \frac{d^2{\bf r'}}{\pi R_*^2} I(|{\bf r}-{\bf r'}|,R) \, ,
\label{eqn:smooth}
\end{equation}
where
\begin{equation}
I(r,R) = \left\{ \begin{array}{ll} 1 + U_1^2(R,r) + U_2^2(R,r) - 2U_1(R,r) \sin \left(\frac{r^2+R^2}{2R_F^2}\right) \\
\quad +\;  2U_2(R,r) \cos \left(\frac{r^2+R^2}{2R_F^2}\right)\; \, \quad \quad \quad r \geq R  \\
U_0^2(r,R) + U_1^2(r,R)\; \quad \quad \quad \quad \quad \quad r < R  \, .
\end{array} \right.
\end{equation}
The functions $U_n(x,y)$ are Lommel functions:
\begin{equation}
U_n(x,y) = \sum_{k=0}^{\infty} (-1)^k \left(\frac{x}{y}\right)^{n+2k} J_{n+2k}\left(\frac{x y}{R_F^2}\right) \, .
\end{equation}
Compared to the geometric description,
the diffraction fringes separate $R$, $D$ and $v$, and is no longer restricted to combinations of $R/v$ and $D/v$.
Additionally, $R_*$  can also be estimated through associated smoothing though to improve the determination of other three parameters,
prior knowledge on $R_*$ is preferred.

\begin{figure}[t]
\centerline{\psfig{file=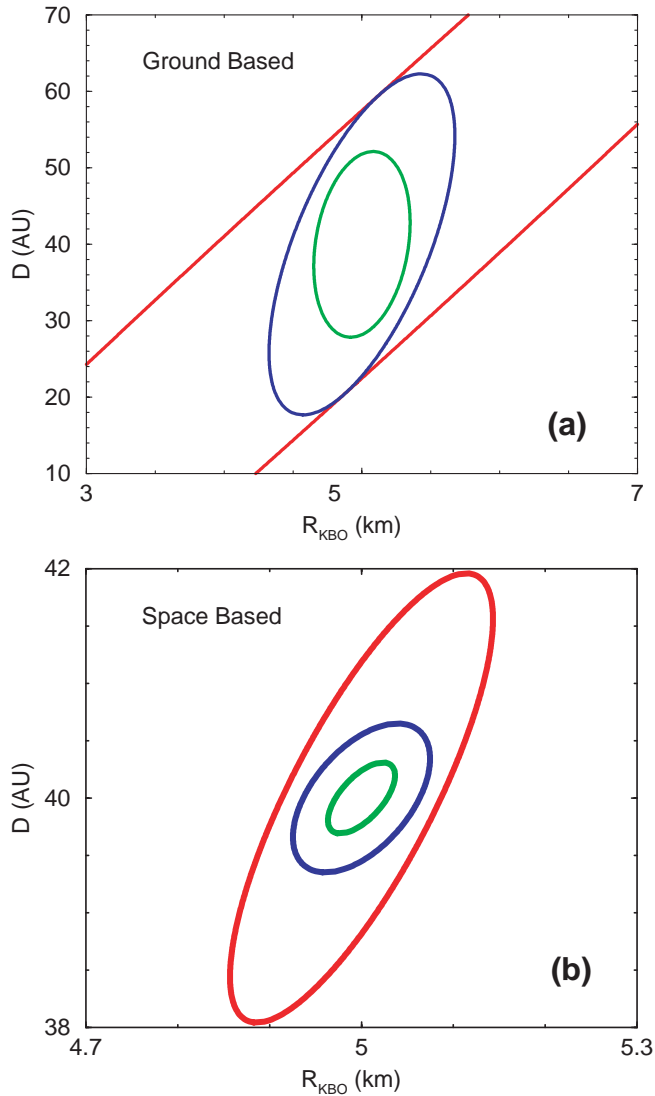,width=3.4in,angle=0}}
\caption{Constraints on the KBO radius and distance, assumed to be at
40 AU with a radius of 5 km, based on the occultation light curve. In (a), we consider
a ground based survey with 0.1 sec timing capability and normalized flux errors of
10\% at the 12th magnitude corresponding to a stellar projected size of 3 km. 
With no prior information on the stellar angular size, one cannot constraint
any of the KBO  parameters. As shown with parallel lines,
with prior knowledge on the projected star size at an accuracy of 10\%, one constraints
a combination of distance and radius  while the two are not independently determined.
To break this degeneracy, prior information on the relative velocity, say from a telescope
array, is needed and with velocity known
to an accuracy of 5\% (20 $\pm$ 1 km  sec$^{-1}$) this leads to
an useful determination of the radius but not the distance (middle ellipse). 
To further improve, we consider an optimistic case involving a 1\% 
known value on the projected star size and 5\% normalized flux errors (inner ellipse).
In this case, the radius can be determined to a level of 10\% while the distance still remains
uncertain at a level of 25\%. 
In (b), we consider a potential space based scenario with timing at
0.1 seconds and a precisely known wavelength for observations. 
Note the  change in y-axis range between (a) and (b).
With a star of projected size 0.1 km, known to 5\% accuracy apriori, and with normalized flux
errors at 5\%, one breaks the degeneracy in KBO distance and size
such that the radius is extracted at a level better than 2\% and the
distance is determined at a level of 5\% (outer ellipse).
Further improvements are obtained when the star size is known more precisely:
middle ellipse shows the case with 1\% while the inner ellipse
shows an optimistic case with a precisely known value for the projected size of the background star
 and 1\% error on normalized flux.}
\label{fig:constraint}
\end{figure}

\section{Practical Aspects}

\subsection{Parameter Extraction}
\label{sec:parameter}

When considering a potential search for KBOs based on occultation observations and to understand
relative merits of a ground vs. a space based survey, we first consider
the ability to extract physical properties of the KBO population from individual
occultation light curves. For this purpose, we make use of the {\it Fisher information matrix}:
\begin{equation}
{\bf F}_{ij} = -\left< \partial^2 \ln L \over \partial p_i \partial p_j
 \right>_{\bf x} \, ,
\label{eqn:likelihood}
\end{equation}
whose inverse provides the optimistic covariance matrix
for errors on the associated parameters (e.g., Tegmark et al. 1997).
In Eq.~\ref{eqn:likelihood}, $L$ is the likelihood of observing data set ${\bf x}$, in our case
the light curve, given the parameters $p_1 \ldots p_n$ that describe these data.
Following the Cram\'er-Rao inequality (Kendall \& Stuart 1969),
no unbiased method can measure the {\it i}th parameter
with standard deviation  less than $({\bf F}_{ii})^{-1/2}$ if all other parameters
are exactly known, and less than $[({\bf F^{-1}})_{ii}]^{1/2}$, from the inverse of the Fisher matrix,
 if other parameters are to be estimated from the  data as well. 

For the present case involving KBO occultation observations, 
we assume light curves will be sampled with time intervals of 
$\delta T$ $(< \Delta T)$. 
To describe a real observational situation, we also assume noise in the light curve
with a variance, $\sigma^2(A_{\rm diff})$, consistent with the instrumental description.
We can now write the Fisher matrix associated with KBO occultation light curves as
\begin{equation}
{\bf F}_{ij} = \sum_{\delta T} \frac{1}{\sigma^2(A_{\rm diff})} \frac{\partial A_{\rm diff}}{\partial p_i} \frac{\partial A_{\rm diff}}{\partial p_j} \,.
\label{eqn:Fisher}
\end{equation}
Here, $\partial A_{\rm diff}/\partial p_i$ denotes derivatives of the light
curve, $A_{\rm diff}$ (Eq.~\ref{eqn:smooth}),  with respect to the
parameter $p_i$.    

In Fig.~\ref{fig:diff}, we show light curves and their derivatives with respect to the four parameters. 
In order to show the importance of diffraction in breaking degeneracies, we show these derivatives
as a function of the projected stellar size. As this size is increased, diffractions fringes 
begin to disappear and the derivatives with respect to physical parameters become smaller such that
the ability to extract parameters from the light curve is reduced. Similarly, with increasing
projected size for the background star, 
the light curves resemble that of a geometric occultation and the discussion
related to the presence of degeneracies apply.  

To show this explicitly, we now consider parameter extraction and
illustrate in terms of the KBO distance and radius, by marginalizing over other two parameters involving
the projected star size and the relative velocity. 
In the case of bright star occultations, as expected from a ground based survey,
there is a significant degeneracy such that one cannot establish the size independent of its distance.
We illustrate this in Fig.~\ref{fig:constraint}(a) 
with the two outer lines. To break this geometric degeneracy,
one requires prior information related to the relative velocity during the event. 
In the case of ground based surveys, this information must come from a second observation of
the same event along the shadow path as it moves across the observer plane. 
Assuming few kilometer baselines, the best one
can hope to achieve from ground, especially for slow occultations towards quadrature, is at the level of a 1 km sec$^{-1}$. 
Using this value as a prior,  one can determine the radius to
an accuracy of 10\%, while the distance still remains uncertain at a level of 50\%.
This can be understood due to the fact that the uncertainty in the projected star size
degrades the distance determination. To further improve, one should have precise knowledge on
the projected sizes of background stars. The most significant improvements are associated with smaller, and thus fainter, 
stars such that diffraction effects become significant.

In Fig.~\ref{fig:constraint}(b), we illustrate such a possibility using the same KBO parameters 
but now with a projected size for the background star of 0.1 km. This size
roughly corresponds to peak of the main sequence stars with  V-band magnitudes between
14 and 16 (for example, Fig.~1 in Cooray \& Farmer 2003).  
Now, one breaks degeneracies significantly such that the radius is known to better than 2\% and
distance is known to 5\%.  Note that this was achieved with no knowledge on the relative velocity.
The diffraction fringes allow the relative velocity to be determined together with
KBO size and distance. We have assumed a single wavelength observation here.
In a real situation, a finite width over wavelength is expected and this  will lead to an additional smoothing of
fringes. Assuming a 25\% width in wavelength of observations around 500 nm,
we determined that errors on KBO size and distance are increased by a factor of 1.5 to 2 from the case
with a single wavelength. This is under the assumption that the filter shape is known precisely.
If the filter function is treated as an unknown, a few percent
constraint on any of the physical parameters is not possible.

\subsection{Surveys}

In prior discussions (Roques \& Moncuquet  2000; Cooray \& Farmer 2003), the consideration  was given to
the ability of KBO occultation
surveys to constrain the KBO size distribution through statistics related to time duration and
normalized flux change.  Instead of this statistical approach, we 
have now considered the possibility of using individual light curves to extract 
physical parameters of the KBO population.
Unfortunately, as discussed, there are significant limitations
on the ability of ground based observations to measure KBO size independent of its distance. To break this geometric degeneracy,
an array of telescopes, with capability to perform relative timing between individual
elements, can be used for simultaneous observations and to measure the relative velocity associated with detected events.
Since one expects KBOs with sizes at the level of
few kilometers to be detected from ground based arrays, individual telescopes should be separated at the same size.
Though we have not considered in detail, there is an interesting optimization problem related to such an array: 
given the fixed number of telescopes, one can design the layout that will
return the most information related to the KBO population. A simple estimate suggests circular arrays with
separations both smaller  and larger  than the typical KBO size one expects to detect; the latter is to reduce
systematics, as in the approach related to TAOS, while the former allows a relative velocity measurement.
Due to small separations of order few hundreds of meters or below, it is unlikely that TAOS will be able to make reliable measurements of
relative velocity associated with its occultation detections. For such surveys, information
related to the KBO population can be extracted through statistics related to two 
observables involving the event duration and the normalized flux variation.

Beyond TAOS, it will be useful to have an optimized array of small telescopes for occultation observations, though
the information on the KBO population one extracts is still limited. 
We demonstrated this based on a calculation related to
the parameter extraction and showed that while KBO radii may be determined at the level of 10\% or better,
KBO distances will remain significantly uncertain. The significant improvements come with observations of
occultations by KBOs of fainter stars, V-band magnitudes fainter than 14, such that diffraction effects are significant.
The advantage here is that no separate relative velocity measurement is needed. For a successful ground based survey that
attempts to make use of diffraction, observations of fainter stars must be achieved with relative flux measurements
down to few percent with timing capabilities around 0.1 seconds.

To further improve, a dedicated space based satellite mission is helpful
with the ability to monitor 14th magnitude or fainter
stars with relative flux errors below 1\% in integration times of order 0.1
seconds or better. The planned Kepler mission to detect transits of extra-solar planets across their central
stars provide a basis to compare what will be required for KBO occultation observations.
The Kepler mission allows the detection of relative flux variations  at the
level of $10^{-5}$ on time scales of 6 hours when 12th magnitude dwarf stars are involved.
Scaling their expectations, we find that in an integration time of 0.1 seconds,
at the 14th magnitude, the photon noise will lead
to Poisson noise contributions to relative flux measurements of order 1\%, which is the level required
for a KBO survey. Since the probability for occultations is relatively high, the data rate can be controlled by a priori
selecting few hundred to thousand stars for monitoring purposes.

To conclude, considering all possibilities, we suggest that a dedicated  mission for KBO occultation
observations may provide the best option to understand the KBO population at
kilometer sizes and below. Unlike observations of reflected light, occultations allow
detections of KBOs, in principle, independent of its distance. The expected large number of
KBOs at the low end of its size distribution leads to a significantly high occultation event rate 
and motivates ground-based attempts for KBO searches. While light curves are likely to be affected 
by a geometric  degeneracy, diffraction provides a useful approach to break this degeneracy and improve
KBO size and distance measurements. The planned TAOS survey provides an initial step in KBO searches 
while significant improvements can be expected with a well planned space-based mission.

\smallskip
{\it Acknowledgments:} 
We thank A. Farmer for useful discussions and acknowledge
partial support from the Sherman Fairchild Foundation.

\newpage

\end{document}